\documentclass[nofootinbib,aps,prl,12pt]{revtex4}
\usepackage[a4paper,margin=1in]{geometry}
\usepackage{float}
\usepackage{newtxtext}
\usepackage[pdftex]{graphicx}
\usepackage{amssymb}
\usepackage{amsmath}
\usepackage{newtxtext}
\usepackage{newtxmath}
\usepackage{bm}
\usepackage{ragged2e}

\begin{document}
\justifying
\setcounter{page}{1}
\title[]{$\boldsymbol{\mathrm{{\mathring{A}}}}$ngstrom-scale ionic streaming when electrical double-layer concept fails}

\author{Jiajia Lu\textsuperscript{1}}
\author{Shuyong Luan\textsuperscript{2}}
\author{Shenghui Guo\textsuperscript{1}}
\author{Libing Duan\textsuperscript{1}}
\author{Guanghua Du\textsuperscript{3}}
\author{Yanbo Xie\textsuperscript{2}}
\email{ybxie@nwpu.edu.cn}

\affiliation{\textsuperscript{1}School of Physical Science and Technology, Northwestern Polytechnical University, Xi’an 710072, China}
\affiliation{\textsuperscript{2}National Key Laboratory of Aircraft Configuration Design, School of Aeronautics and Institute of Extreme Mechanics, Northwestern Polytechnical University, Xi’an 710072, China}
\affiliation{\textsuperscript{3}Institute of Modern Physics, Chinese Academy of Sciences,
Lanzhou, 730000, China}

\begin{abstract}
A knowledge gap exists for flows and transport phenomena at the $\mathrm{\mathring{A}}$ngstrom scale when the Poisson-Nernst-Planck equation based on the concept of electrical double layer (EDL) fails. We discovered that streaming conductance becomes pressure-dependent in $\mathrm{\mathring{A}}$ngstrom channels using latent track membranes. The streaming current emerges only when the applied pressure exceeds a threshold value, which is inconsistent with the existing knowledge as a constant. With increasing channel size, we found that the pressure-dependent streaming conductance phenomenon weakens and vanishes into a constant streaming conductance regime when the mean channel radius exceeds $\sim$ 2 nm. 
The effective surface potential derived from the stream conductance that divides conduction anomalously increases as the channel narrows.
We suspect the pressure-dependent streaming current is due to the reinforced Coulomb interaction between counterions and deprotonated carboxyl groups at the surface, which is close to the ion channel but different from the electrified 2D materials. 
The streaming current emerged due to hydrodynamic friction when the counterions were released from the surface. We approximated the stochastic process of counterion dissociation by 1D Kramer’s escape theory framework and defined the Damk$\ddot{\mathrm{o}}$hler Number to describe the transition from non-linear streaming conductance regime to linear regime as functions of applied pressure and channel radius and well explained the enhanced effective surface potential in confinement. 
\end{abstract}
\maketitle

\section{Introduction}

The flow of water through an electrified surface generates a net ionic current known as streaming current \citep{van2005streaming,bocquet2010nanofluidics,xu2024energy}, which is one of the fundamental electrokinetic phenomena. The study of streaming current is useful in both fundamental research, such as zeta potential characterization \citep{zimmermann2001electrokinetic,werner2001streaming,van2006charge,delgado2007measurement,storey2012effects,siria2013giant,emmerich_enhanced_2022} and applications like energy conversions \citep{olthuis2005energy,xue2017water,marcotte_mechanically_2020,alkhadra2022electrochemical}. Downscaling the channel size benefits the energy conversion by ion-layering effects \citep{gillespie2011efficiently,gillespie2012high} and enhances the efficiency up to 50\% in Nafion systems \citep{catalano2014influence,ostedgaard2017membrane} and power density to $\sim25$ $\mathrm{mW/m^2}$ in Mxene membranes \citep{yang2022simultaneous}.

The knowledge of electrokinetic phenomena and based electrical double layer theory is still developing. 
For instance, ion-ion interactions, typically ignored in classical double layer theories, become critical in high concentrations or divalent electrolyte solutions, named the ionic correlation effect, which precisely describes the electrokinetic phenomena \citep{storey2012effects,mceldrew2018theory,hartkamp2018measuring}. 
Besides, considering the dynamic response of the ions under an external field, the description of electrokinetic phenomena is more accurate than the static double layer theories \citep{bazant2004induced,bazant2010induced}.

However, the concept of the double layer fails when the critical size of the channel comes to the regime of a single hydrated ion, where only a few water molecules exist in the cross-section of the channel. 
New technologies and emerging 2D materials enable us to study flows and mass transport at the molecular level \citep{wang2017fundamental,faucher_critical_2019,aluru2023fluids}. Numerical findings demonstrated that the flow and transport phenomena were inconsistent with the classical understandings such as ultrafast water flows \citep{holt2006fast,radha_molecular_2016,keerthi2021water,itoh2022ultrafast}, voltage-induced mobility change \citep{esfandiar_size_2017}, anomalous dielectric constant \citep{fumagalli_anomalously_2018}, ionic Coulomb blockade \citep{feng_observation_2016, kavokine_ionic_2019,cao2024proton}, breakdown of the Nernst-Einstein relation \citep{li2023breakdown}, quantum friction \citep{kavokine2022fluctuation} and fluidic memristors \citep{robin_modeling_2021,robin2023long,shi2023ultralow}, indicating underlying new physics still need to be unveiled.

Compared to voltage-driven ionic transport, the streaming current in $\mathrm{\mathring{A}}$ngstrom channels was rarely studied in experiments since previous work was conducted on 2D materials that were either charge-free \citep{zhou2023field} or homogeneously charged by voltage gating \citep{marcotte_mechanically_2020} or plasma activation \citep{emmerich_enhanced_2022}. \citet{mouterde_molecular_2019} successfully enhanced the molecular streaming current up to 20 times by voltage gating on carbon nanotubes. Their results showed the ionic mobility derived from the streaming current shows strong nonlinear effects as gating voltage rises, opening a new route to control the ion transport.  
In contrast to non-gated channels, where the streaming conductance grows linearly with applied pressure, voltage-gating induces a Péclet-dependent conductance that results in a nonlinear streaming current \citep{marcotte_mechanically_2020}. \citet{keerthi2021water} studied the molecular streaming in $\mathrm{\mathring{A}}$ngstrom slits and discovered the massive difference in slip length between the graphene and hNB surfaces, unveiling that the electrokinetic mobility of ions increases with the salt concentration and the slipping surface. 

However, for most dielectric materials, such as biological ion channels, the surface acquires charges due to dissociation or adsorption of protons or charged species at functional groups \citep{behrens2001charge,noskov2004control,li2019fast,tan2020hydrophilic}. 
Each charged site bears an elementary charge at the dielectric material surface, which is different from the homogeneous charge distribution at the surface of metal-like graphene or carbon nanotube, where charge distribution may significantly impact the flow \citep{xie2020liquid} and ionic transport in confinement \citep{kavokine_ionic_2019,xie_surface-charge_2023}. 
Thus, the electrokinetic phenomena are still unclear in $\mathrm{\mathring{A}}$ngstrom channels when the double layer concept failed. 

In this work, we investigated the streaming conductance of charged latent track $\mathrm{\mathring{A}}$ngstrom channels. 
We found that in $\mathrm{\mathring{A}}$ngstrom-scale latent track channels, the streaming current emerges when the applied pressure exceeds a threshold value and represents a strong nonlinear rising feature as pressure increases, inconsistent with the existing understanding of the streaming conductance known as a constant \citep{van2005streaming}. However, as the diameter increased, the nonlinear feature of streaming conductance gradually weakened and eventually vanished, becoming pressure-independent as described by classical nanofluidic theories. We estimated the surface potential of nanochannels by streaming conductance divided by conduction, which anomalously increases as the channel narrows when external pressure is applied to the membrane. We attributed the pressure-dependent streaming conductance to the presence of counterions bound to the surface charge when channels come to the $\mathrm{\mathring{A}}$ngstrom scale. The rise in applied pressure increases the hydrodynamic friction on the counterions, increasing the probability of counterions dissociating from the surface charge and thus forming the streaming current. As the channel size increases, the Coulomb interaction between counterions and surface charge reduces, resulting in constant streaming conductance again. We approximated the pressure-dependent streaming conductance by using 1D Kramer's escape approach, which qualitatively explained the experimental observation in $\mathrm{\mathring{A}}$ngstrom channels, including the impacts of applied pressure and size effects. With the defined Damk$\ddot{\mathrm{o}}$hler number, we could identify the pressure-dependent and independent regimes by a phase diagram of the dimensionless coefficient $Da/(Da+1)$. Our work provides anomalous streaming current phenomena in $\mathrm{\mathring{A}}$ngstrom channels that show new phenomena into flows and ionic transport at the smallest scale, which is possibly useful for ion separations or energy applications in the future. \\


\section{Systems and Characterization}

\begin{figure}
\centering
\includegraphics[width=14 cm]{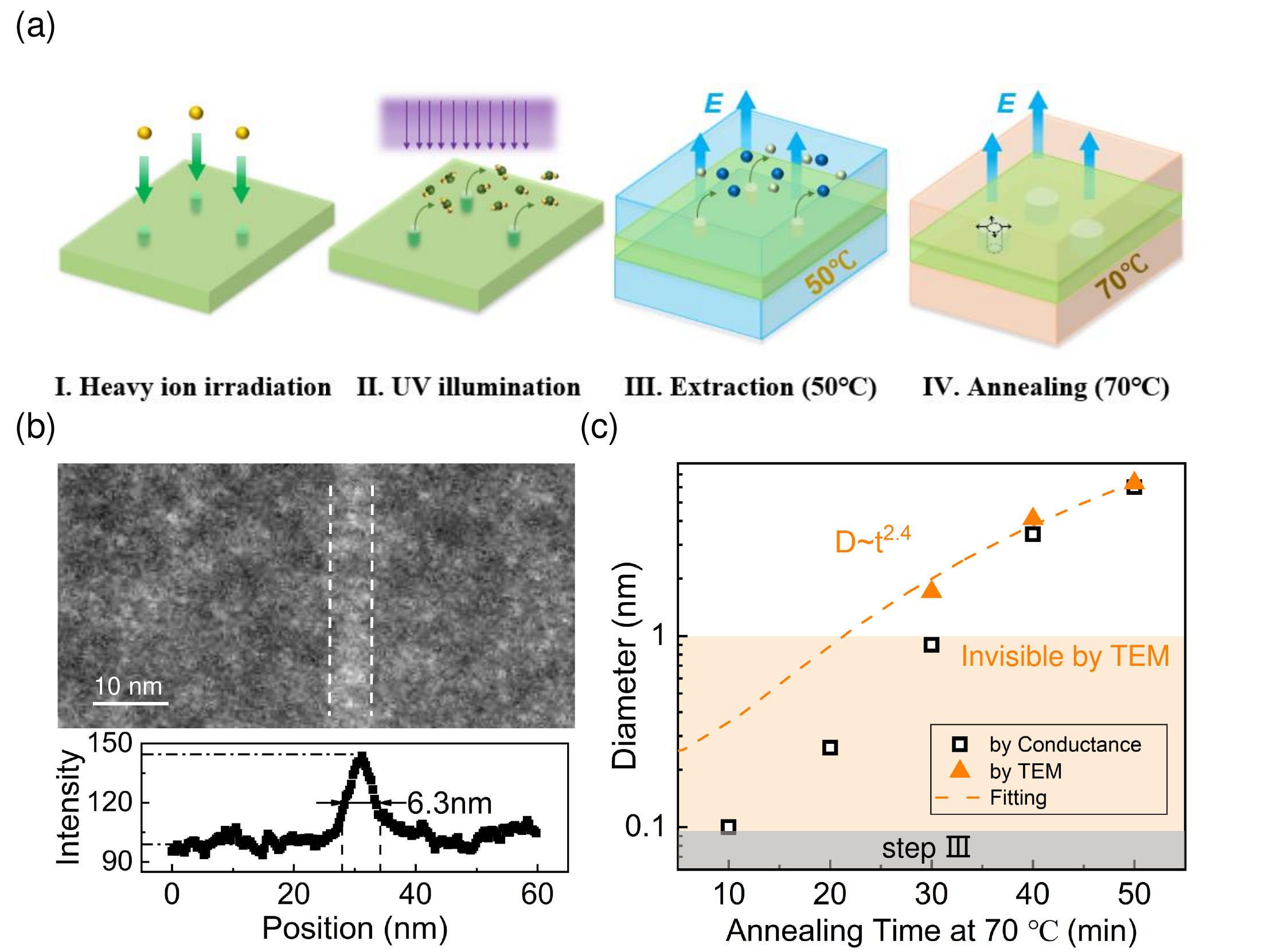}
\caption{(a) The fabrication process of latent track channels, including heavy ion irradiation (step I), UV irradiation (step II), electrophoresis to remove the radiolysis products in a 50$^{\circ}\mathrm{C}$ water bath (step III), and thermal annealing, if needed, in a 70$^{\circ}\mathrm{C}$ water bath to increase the channel size (step IV). The schemes are not to scale. (b) The latent track channels stained by $\mathrm{RuO_4}$ are represented as bright straight lines under TEM. We statistically measured the brightness projected onto the bottom line and took the half peak width of the brightness as the channel diameter. (c) The radius calculated from the measured conductance (black square dots), increases with annealing time. However, only channels with diameters larger than 2 nm are visible under TEM (yellow triangle dots). A numerical fit (yellow dashed line) provides a reference for the channel size. Conductance measurements may underestimate channel size due to blockade effects in confinement.
}
\label{fig1}
\end{figure}

Fig. \ref{fig1}a illustrates the fabrication process of latent track membranes \citep{wen_highly_2016}. 
The $12$ $\upmu\mathrm{m}$ thick PET (Polyethylene terephthalate) foil was first irradiated by $\mathrm{Kr^+}$ ion with energy of $6$ $\mathrm{Mev/u}$ with beam density of $3\times 10^8$ $\mathrm{/cm^2}$ from Lanzhou Heavy Ion Research Facility (HIRFL). Then, we irradiated both sides of the specimens under UV light ($65$ $\mathrm{mW/cm^2}$) for $10$ min to decompose the damaged zone along the latent track \citep{schauries_structure_2018}.
The photo-decomposed polymer molecules could be electrophoretically removed using $0.01$ $\mathrm{mol/L}$ $\mathrm{KCl}$ solution at 50$^{\circ}\mathrm{C}$ at least half an hour, named soft etching \citep{wen_highly_2016,apel_soft-etched_2022,guo_efficient_2024}. Once the conductance reached a saturated value, we believe all radiolysis products have been removed from the channels, resulting in the formation of 1D $\mathrm{\mathring{A}}$ngstrom-scale channels as we characterized by isotherm adsorptions \citep{xie_surface-charge_2023}. The membranes were well cleaned and preserved in a thermostat for overnight.

To study the size effects of nanochannels, we developed the channel size by annealing latent track channels in a water bath at 70$^{\circ}\mathrm{C}$, similar to the method reported by \citet{schauries_structure_2018} in air. Annealing enlarges the size of polymer nanochannels, caused by the damaged track halo diffusing into the track core \citep{schauries_structure_2018}. The naming rule of specimens is as follows: $'\mathrm{TA}'$ denotes Thermal Annealing, with the subsequent number indicating the temperature of the water bath (e.g., $'\mathrm{70}'$ represents 70$^{\circ}\mathrm{C}$). The number following the en-dash ($'\mathrm{-}'$) indicates the annealing time at the given temperature in minutes. The asterisk $'*'$ indicates an additional specimen under the same fabrication process. For example, $'\mathrm{TA70-10}'$ refers to a specimen thermally annealed at 70$^{\circ}\mathrm{C}$ for 10 minutes and $'\mathrm{TA70-10^*}'$ represents the additional specimens thermally annealed at 70°C for 10 minutes. The conductance of each specimen was measured by sweeping voltage in a $0.01$ $\mathrm{mol/l}$ $\mathrm{KCl}$ solution at room temperature. However, the radius of $\mathrm{\mathring{A}}$ngstrom channels can't be precisely characterized by the conductance anymore, due to the strong nonlinear features \citep{xie_surface-charge_2023}. 
Here we take the conductance $G$ at $10$ $\mathrm{V}$ for later discussion, as our previous work found the calculated radius get closer to the true radius in $10$ $\mathrm{V}$ or even higher voltage scanning due to the counterion blockade effects.

Fig. \ref{fig1}c shows that the channel diameter characterized by $G$ at $10$ $\mathrm{V}$ through Ohm's law is proportional to the time of annealing, where the gray area indicates the diameter characterized by the conductance after soft etching (step III) but before annealing (step IV). 
We found the I-V curves (Fig. S2) gradually transition from nonlinear to linear features as the channel radius increases. 
Here, we characterized the latent track nanochannels using ruthenium tetraoxide staining under TEM \citep{adla2003characterization}.
Each specimen was exposed to the vapor of $\mathrm{RuO_4}$ solvent over 20 minutes to guarantee their reactions with unsaturated functional groups. 
We embedded membranes in epoxy resin and sliced them into $100$ $\mathrm{nm}$ sections by an ultrathin slicer $\mathrm{LEICA-UC7FC7}$. The nanochannels could be recognized as a straight, bright line under electron microscopy, as shown in Fig. \ref{fig1}b, due to the ruthenium dioxide deposition. 
We counted the average brightness of the stained region using the software "DigitalMicrograph". The averaged diameter of $\mathrm{\mathring{A}}$ngstrom channels based on the width at half maximum of the TEM is shown in Fig. \ref{fig1}b. The statistical results of the annealed specimen $\mathrm{TA70-50}$, which stands for soft etching in 50$^{\circ}\mathrm{C}$ water bath and Thermal-Annealing at 70$^{\circ}\mathrm{C}$ for 50 minutes, showed that the typical width of the channel is $6.3$ $\mathrm{nm}$, which is exactly the same as we derived from the conductance of $6.1$ $\mathrm{nm}$, shown as orange dots in Fig. \ref{fig1}c.

However, as the channel size decreases to $\mathrm{\mathring{A}}$ngstrom channels, the brightness contrast in the stained area becomes apparently lower due to much fewer unsaturated functional groups. 
We counted the half maximum of the brightness in TEM images, which is proportional to the amount of $\mathrm{Ru}$ molecules penetrating into the polymers, to predict the channel size as a dashed line shown in Fig. \ref{fig1}c (more details are provided in Fig. S1). 
We numerically fitted the diameter change as a function of annealing time $t$ at 70$^{\circ}\mathrm{C}$ water bath as $D \sim t^{2.4}$. The TEM-characterized diameter ($D$) matched well to the calculated ones from conductance when $D>2$ $\mathrm{nm}$ in the absence of non-linear conduction. 
 
\begin{table}
  \begin{center}
\def~{\hphantom{0}}

  \begin{tabular}{c c c c c}  
 \hline\\[3pt]
      \textbf{Specimens} &    
      
      \textbf{\parbox{4cm}{\centering Diameter from Conductance \\ (square dots)}} & 
      \multicolumn{2}{c}{\textbf{\parbox{3.8cm}{\centering Diameter by TEM \\ (Triangular dots)}}} & 
      \textbf{\parbox{3.5cm}{\centering Diameter from Fitting \\ (dashed line)}} \\[5pt] 
    
      TA50      & 0.05 nm   & \multicolumn{2}{c}{Invisible} & 0.2 nm \\
      TA70-10   & 0.1 nm    & \multicolumn{2}{c}{Invisible} & 0.4 nm \\
      TA70-20   & 0.26 nm   & \multicolumn{2}{c}{Invisible} & 0.9 nm \\
      TA70-30   & 0.9 nm    & \multicolumn{2}{c}{1.7 nm (Visible)} & 2.0 nm   \\
      TA70-40   & 3.4 nm    & \multicolumn{2}{c}{4.1 nm (Visible)} & 3.8 nm \\
      TA70-50   & 6.0 nm    & \multicolumn{2}{c}{6.3 nm (Visible)} & 6.4 nm \\
      \\
      \hline
  \end{tabular}
  \caption{Diameter of channels from calculated conductance, stained specimens under TEM and numerical fitting, respectively.}
  \label{tab:diameters}
  \end{center}
\end{table}
But the diameter derived from conductance starts to deviate from the dashed lines for smaller channels. 
The diameter calculated from the conductance is underestimated, since we found the voltage-activated conductance in latent track channels \citep{xie_surface-charge_2023} with the hypothesis of counterion-blockade that reduces the ionic conductance conduction. The physical interpretation of the numerical fitting remains unclear, as missing of the details like the structures and dynamics of the polymers during thermal annealing. But, at least, the fitting provides a reference value for estimating the channel size, particularly in cases where both conduction measurements and TEM fail to characterize channels with diameters smaller than 2 nm. Table 1 summarizes the sizes characterized using different approaches, providing a basis for analyzing streaming conductance and comparing between the different methods.


\section{Results and discussions}

\begin{figure}[H]
\flushleft
\centerline{\includegraphics[width=14 cm]{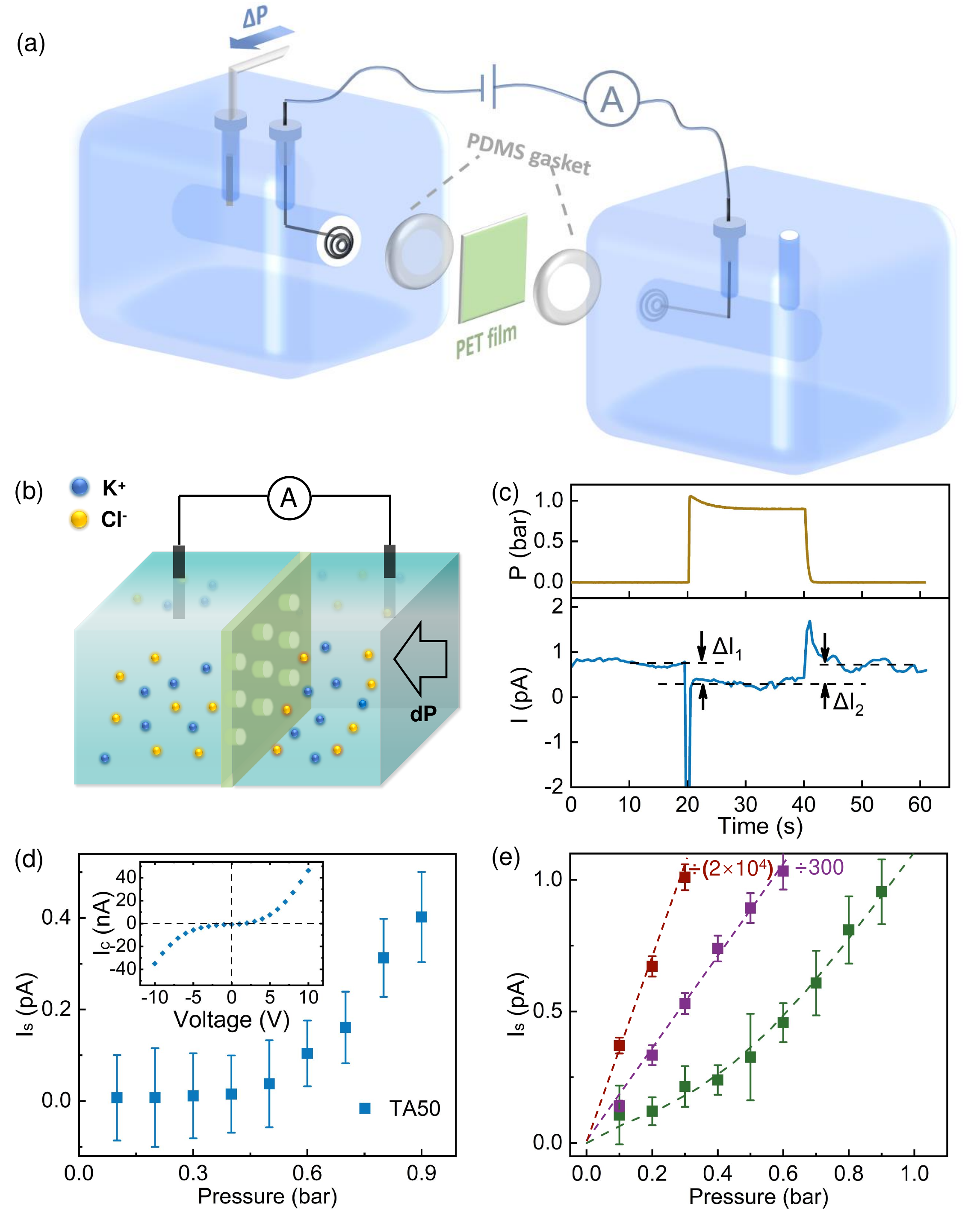}}
\caption{(a) Schematic diagram of the device for streaming current measurement. (b) Schematic description of ion transport across PET membrane under an external pressure. (c) The current recordings (bottom figure) when applied pressure (upper figure) was turned on. (d) The averaged streaming current exponentially increases as applied pressure in specimen TA50, with typical I-V curves in the inset figure that also show a strong non-linear feature. (e) The streaming current gradually turns to linear response to the applied pressure as the channel radius grows. The current amplitudes of specimens TA70-40 and CE were divided by a factor of 300 and 2 $\times$ $10^4$ respectively to make a comparison of the $\Delta p-I_s$ curves over a wide range of channel size.} 
\label{fig2}
\end{figure}

We sealed the left PMMA reservoir and clamped a piece of latent track membrane using two PDMS gaskets, where an approximate 0.8 $\mathrm{cm^2}$ surface area of the membrane was exposed to the electrolyte solution, so that the applied pressure could be controlled from the left reservoir using a pressure pump (Fluigent) (Fig. \ref{fig2}a). Two $\mathrm{Ag/AgCl}$ electrodes were inserted in the reservoirs and connected to a Sub-femtoamp Remote SourceMeter (Keithley 6430) for the current recording. The silver wires with a diameter of 0.6 mm and a length of 15 cm were polished with sandpaper for a smooth surface and rinsed with DI water, and ultrasonically cleaned in ethanol and acetone solutions for 15 minutes to remove the possible organic contamination. The cleaned silver wires were immersed in 0.1 M $\mathrm{FeCl_3}$ for 15 minutes to produce $\mathrm{Ag/AgCl}$ electrodes. The $\mathrm{Ag/AgCl}$ electrodes area exposed to the salt solutions is 1.9 $\mathrm{cm^2}$ approximately and shows a few millivolts difference. We filled reservoirs with $0.01$ $\mathrm{mol/L}$ $\mathrm{KCl}$ solution at $\mathrm{pH = 5.4}$, where the Debye length is $\sim$ 3 nm that is much larger than the channel radius.  
A Faraday cage covered the system to reduce the noise level.
We measured the streaming current from $\mathrm{TA50}$ 
fabricated by soft etching and free of annealing, which has the smallest conduction among all specimens. 

The typical I-V curve of the specimen $\mathrm{TA50}$ in the inset figure of Fig. \ref{fig2}d represents an obvious nonlinear conduction feature. We attributed the high resistance at low voltage to the strong interaction between counterions and surface charge, which may block ionic transport through the 1D $\mathrm{\mathring{A}}$ngstrom channel \citep{xie_surface-charge_2023}. Increases in applied voltage may increase the probability of counterions releasing from the surface charge, thus opening the channel for conduction. 
Thus, it is quite interesting to measure the streaming current in our latent track channel, which is a 1D charged $\mathrm{\mathring{A}}$ngstrom system, since the streaming current is directly proportional to the surface charge that may support the counterion blockade mechanisms. 
The current recordings in Fig. \ref{fig2}c show an obvious current jump from the current noise once pressure was applied with $\Delta$$p = 0.9$ $\mathrm{bar}$, which vanished when the pressure was turned off.
We take the average value of current change $\Delta {I_1}$ when $\Delta p$ turned-on and $\Delta {I_2}$ when $\Delta p$ turned-off as the streaming current, to avoid the drift of current noise due to the ion concentration polarization.
The fluctuating current after the pressure was turned off was caused by the mechanical vibration. 

The Helmholtz–Smoluchowski (HS) equation as well as the Poisson–Nernst–Planck (PNP) equations describe the streaming current at a charged wall $I_s=\varepsilon_r\varepsilon_0\pi R^2\Phi \Delta p/\eta L$, where $\varepsilon_r$, $\varepsilon_0$, $R$, $\Phi$, $\eta$ and $L$ are, relative permittivity of water, vacuum permittivity, channel radius, surface potential (or zeta potential in continuum regime), viscosity and channel length, as here we take the form of HS equation for simplification. The streaming conductance defined as $S = I_s/\Delta p$ is independent of applied pressure \citep{van2005streaming,yaroshchuk2010interpretation,xue2014tunable}, 
$S \sim \varepsilon_r\varepsilon_0 R^2\Phi/\eta L$.

However, our experimental results of streaming current in latent track $\mathrm{\mathring{A}}$ngstrom channels showed a strong nonlinear feature as $\Delta p$ in Fig. \ref{fig2}d, which remains nearly at zero below 0.5 bar but dramatically increases as applied pressure exceeds 0.5 bar. The error bars were statistically derived from the measured streaming current in a single experiment. It seems that the streaming current can be activated by applying pressure. We suspect the pressure-dependent streaming conductance is correlated with the impacts of the $\mathrm{\mathring{A}}$ngstrom channel and heterogeneous surface charge, since it was rarely discovered even in $\mathrm{\mathring{A}}$ngstrom scale channels, only except for the voltage-gated Péclet-dependent conductance \citep{mouterde_molecular_2019,marcotte_mechanically_2020}.

\begin{figure}[H]
\flushleft
\centerline{\includegraphics[width=14 cm]{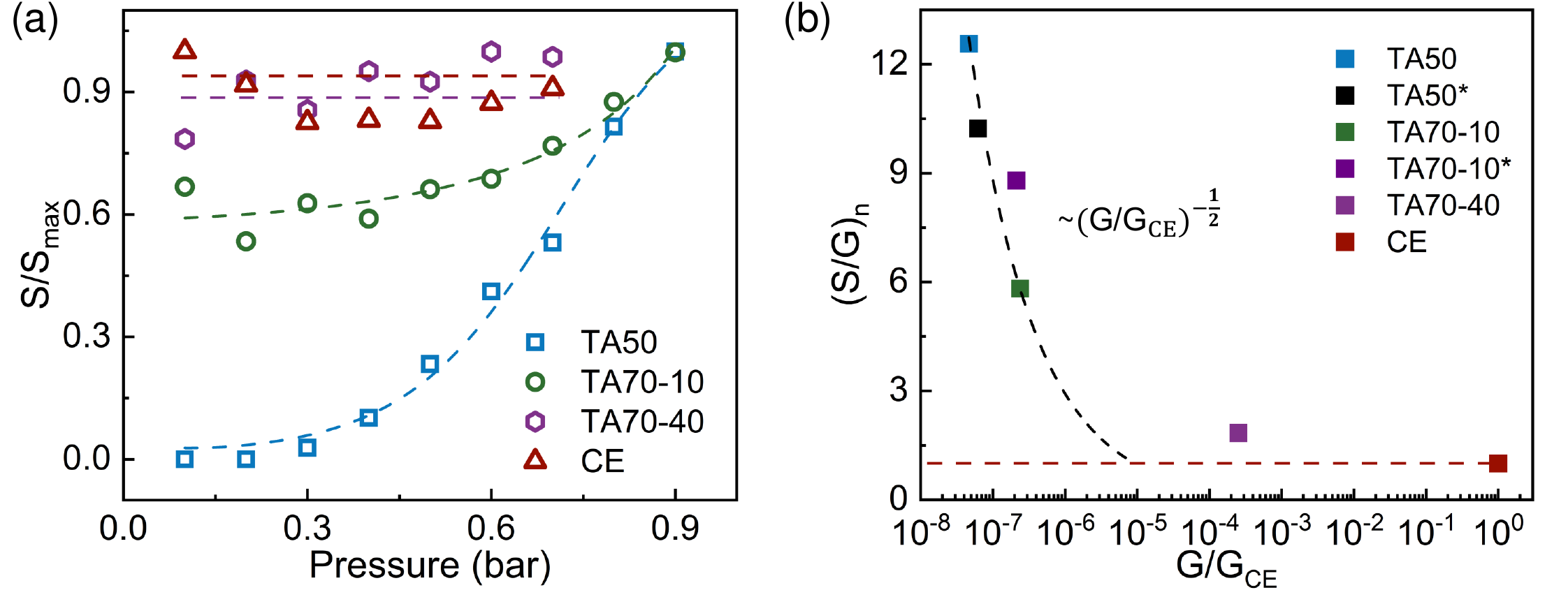}}
\caption{
(a) The normalized streaming conductance ($S/S_{max}$) shows strongly pressure dependency in $\mathrm{\mathring{A}}$ngstrom channels and gradually becomes a constant as radius increases. (b) Normalized $S/G$ versus normalized conductance for specimens across a wide range of radius. For small size channels, the normalized $S/G$ decreases with increasing normalized conductance $(S/G)_n \sim (G/G_{CE})^{-\frac{1}{2}}$, thus reversed proportional to the radius $(S/G)_n \sim R^{-1}$.}
\label{fig3} 
\end{figure}

We investigate the streaming current of annealed or chemically etched membranes with a wide range of diameters.
The mean radius of specimen $\mathrm{TA70-10}$ which was annealed at 70$^{\circ}\mathrm{C}$ for 10 min, is $0.4$ $\mathrm{nm}$ derived from extrapolation of numerical fitting by TEM characterization. Since the streaming current amplitude is proportional to the channel cross-sectional area $I_{s} \sim R^2$, it is challenging to present data spanning a wide range of radii in a small current scale. To better represent the $\Delta p - I_{s}$ curve across a broad range of channel sizes, the streaming current amplitudes for specimen TA70-40 and CE were reduced by factors of 300 and $2\times10^4$, respectively, as shown in Fig. \ref{fig2}e. This normalization enables a direct comparison of the linearity of the $\Delta p - I_{s}$ curve on the same scale. However, the streaming current still represents nonlinear features in specimen $TA70-10$, which is less obvious than specimen $\mathrm{TA50}$. As $D$ increases to $3.4$ $\mathrm{nm}$ estimated from conductance measurement in specimen $\mathrm{TA70-40}$, the nonlinear features disappear in both streaming current and ionic conduction (Fig. S2 in SI). As a control group, we measured the streaming current in specimen $\mathrm{CE}$, which stands for Chemically-Etched with a diameter of $220$ $\mathrm{nm}$ \citep{apel_diode-like_2001,mo_fabrication_2014}. It shows a linear increase in $I_s$ as $\Delta p$ (Fig. \ref{fig2}e) with nearly a constant $S$, represented by purple dots in Fig. \ref{fig3}a. 

To clearly demonstrate the transition of streaming conductance, we first calculated $S=I_{s}/\Delta p$ as a function of applied pressure and normalized the streaming conductance by their maximum value $S_{max}$ at applied pressure from 0.7 bar $\sim$ 0.9 bar as $S/S_{max}$ for the four specimens shown in Fig. \ref{fig3}a. 
For the specimen $\mathrm{TA50}$ with the smallest conductance and size, the $S$ represents an exponential rise in applied pressure that deviates from the existing knowledge of the electrokinetic theory approach. The rest of the $\mathrm{\mathring{A}}$ngstrom-channels also showed such non-linear feature of the streaming current, although in different levels. We found $S/S_{max}$ becomes weaker pressure-dependent as the channel size grows. When the mean radius $R > 2$ $\mathrm{nm}$, we found the streaming conductance becomes independent of the applied pressure in specimens $\mathrm{TA70-40}$ and CE. We could estimate the surface potential $\Phi$ by streaming conductance divided by ionic conduction  $S/G \sim \varepsilon_r\varepsilon_0 \Phi$, since it is still technically challenging to know the $\mathrm{\mathring{A}}$ngstrom channel size from both TEM characterizations and conduction measurement.

Here we normalized $S/G$ by the value derived in specimen $\mathrm{CE}$ (220 $\mathrm{nm}$ diameter), $(S/G)_n=\frac{S}{S_{CE}}\cdot \frac{G_{CE}}{G}$, where $S_{CE}$ and $G_{CE}$ are streaming conductance and conductance of the specimen CE as shown in Fig. \ref{fig3}b. The $(S/G)_n$ is inversely proportional to the square of normalized conductance $G/G_{CE}$ for small channels. According to the Helmholtz-Smoluchowski equation, the surface potential ought to be independent of channel size, however, our results showed that the $\Phi$ increases as the channel narrows, as we discuss in later texts. Additional data with specimens $\mathrm{TA50}$ and $\mathrm{TA70-10}$ are included in Fig. \ref{fig3}b, with the original data found in Fig. S3.

We attributed the pressure-dependent streaming conductance to the strong Coulomb interaction between counterions and the carboxyl group on the surface of latent track channels \citep{wang2018ultrafast}, where the Coulomb interaction can be reinforced in 1D $\mathrm{\mathring{A}}$ngstrom channels according to previous studies \citep{teber_translocation_2005,kavokine_ionic_2019}. Pressure-driven flow within the channels gives hydrodynamic friction to the bound counterions, increasing their probability of releasing surface charge and thus forming the streaming current. We can approximate the streaming current by Kramer's escape problem with an potential well, considering the counterion transport as the summation of a mean first passage time $\tau_{m}$ escaping from the potential well and the time of free drifting $\tau_{d}$. Thus, we have the streaming current as $I_{s}=\frac{N\Sigma A}{\tau_{m}+\tau_{d}}$, where $\Sigma$, $N$, $A$, and $L$ are surface charge density, numbers of parallel channels, inner wall surface area of nanochannels $A=2\pi R L$, and channel length. 
The $\tau_{m}$ can be approximated in a form of $\tau_{m}\sim e^{\Delta U_s-c \Delta p}$, where $\Delta U_s$ is the Coulomb interaction energy normalized by $k_B T$, and $c$ is a friction coefficient for the counterions normalized by $k_B T$. 
At a steady state, the pressure-driven force equals to the friction, $2\pi R\lambda L v_w=\Delta p \cdot \pi R^2$, where $\lambda$ is the friction coefficient including the entrance effects, and $v_w$ is flow velocity. Thus, we have the time of free drifting as $\tau_{d}=2\lambda L^2/(\Delta p R)$. Finally, we have the streaming current in 1D heterogeneously charged $\mathrm{\mathring{A}}$ngstrom channels expressed as follows:

\begin{equation}\label{j}
    j_{s} \sim \frac{2 N L\Sigma/R }{\tau_{m}^0 e^{\Delta U_s-c \Delta p}+2\lambda L^2/(\Delta p R)}
\end{equation}

where $\tau_{m}^0$ is the mean-first passage time in the limit $\Delta U_s \rightarrow 0$. 
Here we could define a Damk$\ddot{\mathrm{o}}$hler 
Number $Da = \tau_{d}/\tau_{m}$ to indicate the ratio between the drifting (residence) time of a counterion transport through the channel $\tau_{d} = 2\lambda L^2/(\Delta p R)$ and the mean first passage time $\tau_{m} =\tau_{m}^0 e^{\Delta U_s-c \Delta p}$ which described the dissociation of counterion from the surface charge that can be considered as a chemical reaction. Thus, we can rewrite the equation of streaming current as follows:

\begin{equation}\label{I_s}
    I_{s} \sim [\frac{Da}{1+Da}\frac{\pi N \Sigma R^2}{\lambda}]\frac{\Delta p}{L}
\end{equation} 

where the factor in square bracket is the Onsager reciprocal coefficient. Thus, we could derive the $S/G$ according to the form of streaming current in form of Helmholtz–Smoluchowski equation:  

\begin{equation}\label{Phi}
    S/G \sim \varepsilon_r\varepsilon_0 \Phi\sim \frac{Da}{1+Da} \cdot \frac{\Sigma \eta}{\lambda} 
\end{equation} 

where $\eta$, $\varepsilon_r$, $\varepsilon_0$ are viscosity, relative permittivity of water and vacuum permittivity. Given the challenges in obtaining details such as Coulomb interactions $\Delta U_s$ and flow velocity in our polymer $\mathrm{\mathring{A}}$ngstrom-scale channels, we calculate the dimensionless coefficient using a unitless contour map as a function of applied pressure and channel radii. We first examine the effects of applied pressure on $Da$ and streaming conductance. For small $Da$ when $\tau_{d}/\tau_{m} << 1$, we have $\frac{Da}{1+Da}\approx Da$, and the streaming conductance is proportional to $Da \sim \frac{e^{-\Delta U_s + c\Delta p}}{R \Delta p}$. This results in an exponential increase of streaming conductance as the applied pressure $S \sim e^{-\Delta U_s+c\Delta p}$ by equation \ref{I_s}. But, as $\Delta p$ exceeds a threshold value, the large $Da$ leading to $\frac{Da}{1+Da} \sim 1$, the streaming conductance reaches a constant value, as shown in Fig. 4c for high $\Delta p$.

\begin{figure}
\flushleft
\centerline{\includegraphics[width=14 cm]{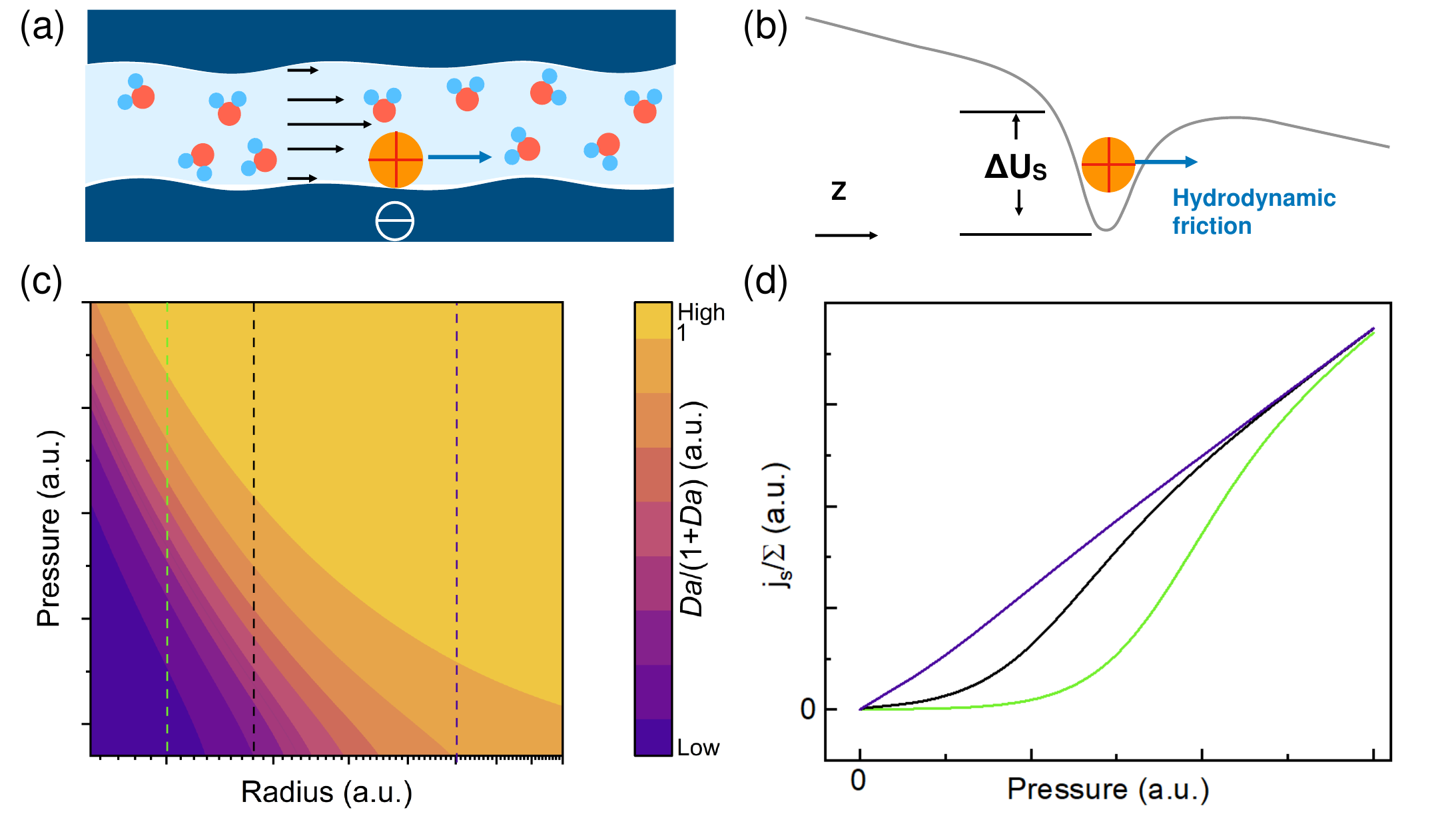}}
\caption{(a) The scheme of molecular streaming with a counterion (yellow) strongly bound to surface charge (white), causing pressure-dependent streaming current. (b) The thermal dynamics process of bound ion under the action of hydrodynamic friction can be described by 1D Kramer’s escape problem from a Coulomb potential well. (c) The dimensionless coefficient $Da/(Da+1)$ describe the nonlinearity of streaming current as functions of applied pressure and channel radius. The yellow region indicates constant streaming conductance zone. (d) The streaming current flux normalized by surface charge density ($j_s/\Sigma$) as a function of applied pressure in three different channel size, where the color corresponds to the dashed lines in Fig. \ref{fig4}c.}
\label{fig4}
\end{figure}

Now we focus on the size effects. As channel radius increases, the Coulomb interaction between counterions and surface decreases $\Delta U_s \sim R^{-1}$ due to the reinforced Coulomb interaction and large Bjerrum length in confinement, as reported in Ref. \citep{teber_translocation_2005,kavokine_ionic_2019}. 
The reduced Coulomb interaction $\Delta U_s$ in large channels increases the $Da$ thus leading to $\frac{Da}{1+Da} \sim 1$ for a linear $I_s-\Delta p$ curve shown as a purple solid line in Fig. \ref{fig4}d. In such case, the streaming conductance becomes a constant. We assume a constant number of binding sites across channels of different sizes, rather than a constant surface charge density, as the binding sites are generated during irradiation instead of the annealing process, even though the channel size gradually increases with annealing. Thereby, we have the surface charge density inversely proportional to $R$, $\Sigma=\frac{ne}{2\pi RL}\sim R^{-1}$, where $n$ is the number of charged sites as a constant. Thus, the $S/G$ in large $Da$ is inversely proportional to the channel radius $(S/G)_n \sim \Sigma \sim R^{-1}$. Previous results \citep{fumagalli_anomalously_2018} showed that the permittivity of water channel $\varepsilon_r$ decreases as the channel radius getting in $\mathrm{\mathring{A}}$ngstrom scale, so we suspect the surface potential $\Phi$ possibly also increases as the channel size decreases, which well explained the anomalous increases of surface potential in $\mathrm{\mathring{A}}$ngstrom channels.

Although directly observing ionic transport and detailing ion-wall interactions remains challenging, we believe our findings provide new insights into electrokinetic phenomena in $\mathrm{\mathring{A}}$ngstrom channels where the concept of electrical double layer fails. 
We found the streaming current can only be activated when applied pressure exceeds a threshold value, which was only observed in a voltage-gated CNT \cite{marcotte_mechanically_2020} although with a different tendency. 

Our results reveal electrokinetic phenomena when the channel size gets close to an atomically thin Helmholtz layer, which also provide evidence of the counterion-correlated blockade of ionic transport in confinement. A generic external force, including electrical force and hydrodynamic friction force, could release the counterions, forming the ionic current. 

Our results with increasing channel radius give a threshold boundary from sub-continuum to continuum transport $\sim$ 2 nm, above which size the ionic transport reaches a regime where flux is linearly proportional to the generic force. In such a sub-continuum regime, when the EDL concept fails, the reciprocal factor between water flow and ionic fluxes needs to be reproduced. For instance, although we derived the reciprocal factor in the form of effective surface potential from streaming current, the fluxes of mass transport are no longer a linear summation of the fluxes driven by individual forces, as the fluxes may exponentially respond to a generic force. When $\Delta U_s/k_B T> 1$, a threshold pressure is required to activate the streaming current, as described by equation \ref{j}. The Coulomb interaction is significantly enhanced in $\mathrm{\mathring{A}}$ngstrom-scale confinement due to the reduced permittivity, which increases the Bjerrum length of water in confinement. When $\Delta U_s/k_B T<1$, thermodynamic energy enables the counterions to dissociate from the surface charge, induces a negligible $\Delta U_s/k_B T$ term, simplifying the equation \ref{j} into the conventional Smoluchowski-Helmholtz equation, where no threshold pressure is needed for streaming current generation. However, the behavior of ions and water near the surface is quite complicated, which requires further understandings of the surface interactions. Our work probably gives information on the Donnan potential of a dielectric surface through an experimental approach to streaming current. It is still challenging to characterize the water flows and mass transport on such a small scale in polymer channels, but our work may help to knowing a bit on the flow and transport in confinement. 
The strong ion-wall interaction will be critical for ionic diffusion and transport in the dielectric $\mathrm{\mathring{A}}$ngstrom channels and may be useful in ionic separation and energy applications.


\section{Conclusions}
We report pressure-dependent streaming conductance in heterogeneously charged latent track \AA ngstrom channels, which contradicts existing knowledge based on the electrical double layer theories that the streaming conductance is pressure-independent. To investigate the size effects, we developed the channel size from the \AA ngstrom scale to a few nanometers by thermal annealing in a water bath and characterized it by staining under TEM. As the channel radius increases above $\sim$ 2 $\mathrm{nm}$, the streaming conductance gradually becomes constant, as predicted by the PNP and Helmholtz-Smoluchowski equations. 

We derived the anomalously effective surface potential increasing as the channel narrows by dividing streaming conductance by conduction, which we attribute to the increase in surface charge density with decreasing channel size in \AA ngstrom channels. Our hypothesis attributed the pressure-dependent streaming conductance to the strong interaction between counterion and surface charge, where the rises of applied pressure increases the friction force on the bound cations and probability of releasing from the surface, thus forming current. The Coulomb interaction decreased with an increase in channel radius, resulting in a constant streaming conductance. 

We quantitatively approximated the ionic transport using the 1D Kramer's escape theory framework, which well explained the pressure-dependent streaming conductance with the impacts of size effects. 
We derived a phase diagram of dimensionless coefficient $Da/(Da+1)$ based on Damk$\ddot{\mathrm{o}}$hler Number, which can distinguish between pressure-dependent and pressure-independent regimes as functions of radius and pressure gradient. 
We discovered pressure-dependent streaming current in heterogeneously charged \AA ngstrom channels, where surfaces are charged via deprotonation or physical adsorption processes for dielectric surfaces. 
Our work is helpful for understanding ionic transport in confinement and may be useful for energy applications and ion separations.
\section{acknowledgments}
\begin{acknowledgments}
The authors thank Prof. Jie Liu and Prof. Jinglai Duan in HIRFL for their help with film irradiation and useful discussion for the $\mathrm{RuO_4}$ stained TEM characterization, and thank Shenghui Guo for the help of TEM characterization.
This work was supported by the National Natural Science Foundation of China No. 12388101, 12241201,12075191, and the Fundamental Research Funds for the Central Universities with Grant No. D5000210626.
\end{acknowledgments}


\bibliography{reference}

\end{document}